\documentclass[a4paper,11pt]{article}

\pdfoutput=1 

\usepackage{jinstpub} 

\usepackage{soul}
\usepackage{color}
\usepackage{multicol} 
\usepackage{array} 
\usepackage{url} 
\usepackage{wasysym}
\usepackage{hyperref}
\usepackage{eurosym}
\usepackage{pifont}
\usepackage{subfigure}
\usepackage[bottom]{footmisc}
\usepackage{lineno}
\usepackage{gensymb}
\title{\boldmath First testing of the Hamamatsu R5912-02Mod photomultiplier tube at 4-bar pressure and cryogenic temperature}

\author{D.~Belver}
\author{E.~Calvo,}
\author[1]{C.~Cuesta,\note{Corresponding author}}
\author{A.~Gallego-Ros,}
\author{I.~Gil-Botella,}
\author{S.~Jim\'{e}nez,}
\author{C.~Lastoria,}
\author{I.~Mart\'{i}n,}
\author{J.J.~Mart\'{i}nez,}
\author{C.~Palomares,}
\author{J.~Soto-Oton,}
\author{A.~Verdugo}

\affiliation{Centro de Investigaciones Energ\'{e}ticas, Medioambientales y Tecnol\'{o}gias (CIEMAT),\\ Madrid, Spain}

\emailAdd{clara.cuesta@ciemat.es}

\abstract{The Hamamatsu R5912-02Mod photomultiplier tube (PMT) will be used in the DUNE dual-phase module, a 10-kton fiducial volume liquid-argon time-projection chamber, which is one of the four projected far-detector modules of the DUNE long-baseline neutrino experiment. In the DUNE dual-phase module, the liquid argon places high pressure on the photo-detectors located at the bottom of the 14-m cryostat. Four Hamamatsu R5912-02Mod PMTs were tested at 4-bar absolute pressure and cryogenic temperature (in liquid nitrogen) for the first time. No mechanical or electrical damage is reported, validating the use of this PMT model in the DUNE dual-phase module and in other large scale cryogenic liquid detectors. The differences observed in their behavior are expected for the change in the PMT operating temperature.}

\arxivnumber{2003.13463} % only if you have one

\begin{document}
\maketitle
\flushbottom

\section{Introduction}
\label{sec1}

In particle physics, large scale cryogenic liquid detectors for neutrino, dark matter, or other rare event searches require photon detection systems able to support the pressure placed by the liquid where the photo-detector is immersed maintaining their performance. A low-cost, large detection area, and efficient option is the use of photomultiplier tubes (PMTs). However, no validation measurements of PMTs in over-pressure and cryogenic conditions are found in the literature, nor carried out by the manufacturer.

In particular, DUNE is a dual-site experiment for neutrino science and nucleon decay searches~\cite{DUNEtdrv1, DUNEtdrv2}. The far detector will consist of four 10-kton fiducial-mass liquid-argon time-projection chambers (LAr TPCs). One of these modules is foreseen to use LAr TPC dual-phase (DP) technology and the design of the photon detection system calls for PMTs distributed uniformly on the bottom of the cryostat~\cite{duneIDRv3}. According to the baseline design, 720 PMTs, approximately 1 per m$^2$, will be installed at the DP module. The PMT signals add precise timing capabilities being essential for a full 3D particle reconstruction and for cosmic background rejection.

In the DUNE DP module, PMTs will operate at 3-bar absolute pressure as they will be placed under 14\,m of LAr. Allowing for a safety margin, we carried out the measurements at 4-bar absolute pressure using the same instrumentation and following the same procedures as for the PMT cryogenic characterization~\cite{protoDUNEPMTs}. The goal of this test was to prove that the selected PMT model supports this pressure in cryogenic conditions without showing mechanical or electrical damage. The manufacturer has validated the PMTs to operate up to 7\,bar at room temperature (RT), but does not perform any test in both cryogenic temperature (CT) and over-pressure. The potential risks are a vacuum loss causing the malfunctioning of the device, and, in the worst case, a glass break with fateful consequences in the detector~\cite{SKreport}. Other experiments \cite{Wang:2017xvc,Abbasi:2010vc} enclose the PMT into a shielding structure to avoid PMT implosion, but the vacuum ultraviolet (VUV) light  from LAr scintillation (128 nm) would be absorbed by such a system.

\section{Hamamatsu R5912-02Mod PMT}
\label{sec2}

The Hamamatsu R5912-20Mod cryogenic PMT\footnote{https://www.hamamatsu.com/}  has a 14-stage dynode chain which provides a nominal gain of 10$^9$ at RT. This PMT model  was successfully operated in the WA105 3$\times$1$\times$1\,m$^3$ demonstrator~\cite{311}, and in the ProtoDUNE-DP detector~\cite{Cuesta:2019yeh, protoDUNEPMTs, Belver:2019lqm}. Also, similar PMTs are used in other particle physicis experiments like MicroBooNE~\cite{microboone1}, MiniCLEAN~\cite{clean2}, ArDM~\cite{ArDM2}, ICARUS T600~\cite{icarus}, and SBND~\cite{Machado:2019oxb}. 

In this work, four PMTs were tested: one PMT (serial number: FB0019) previously characterized at CT in a similar set-up, and three PMTs provided by Hamamatsu for these tests with known defects: high dark current and after-pulse rates (FA0169), peeling of evaporated aluminum on glass bulb (FA0175), and low cathode sensitivity (SN: FA0187). In this last PMT, we observed gain instabilities, so we only used it as a mechanical sample. In the others, the defects did not affect the results.

\section{Cryogenic set-up and methodology}
\label{sec3}

A dedicated test bench is designed for the PMT characte\-rization. 
The CT measurements are performed with the PMT inside a 50\,L vessel filled with liquid nitrogen (LN$_2$).

\begin{figure}[ht]
\includegraphics[width=0.9\textwidth]{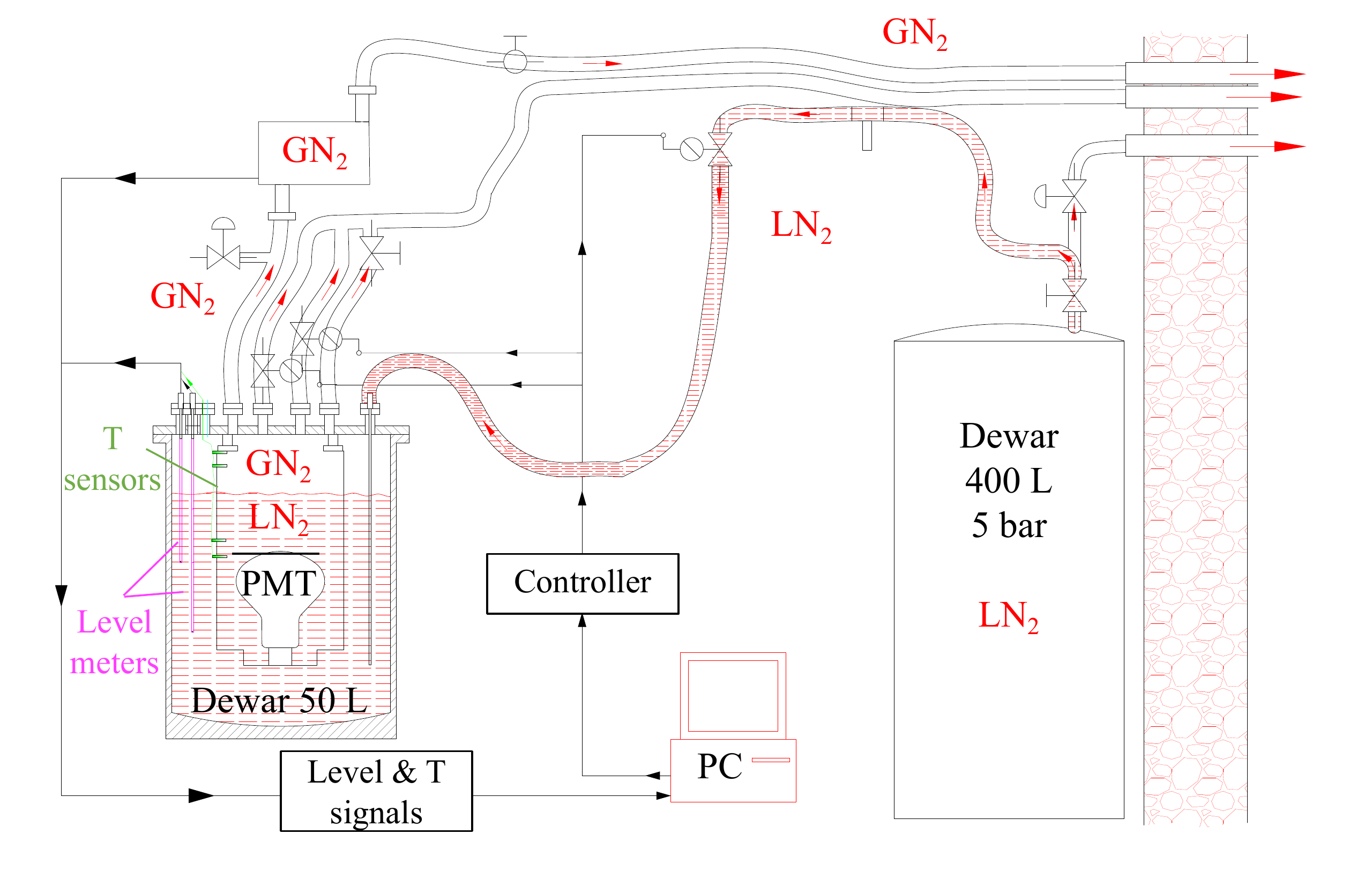}
\centering \caption{Detailed cryogenics set-up for PMT testing up to 4-bar pressure.}
\label{fig:setup2}
\end{figure}

A 400\,L tank supplies LN$_2$ at $\sim$5\,bar pressure through a pipe directly to the 50\,L vessel where the PMT is located. A diagram of this system is shown in Fig.~\ref{fig:setup2}. The system is automatically filled, using electro-valves regulated by level and temperature probes through a PC. The over-pressure is created by  the evaporated N$_2$ inside the 50\,L vessel and controlled through a back-pressure regulator. The cables of the PMTs, temperature and level sensors, and the optical fiber pass through several CF40 ports.

The PMT voltage divider and support structure used in ProtoDUNE-DP~\cite{protoDUNEPMTs} and projected in DUNE are also employed in these tests.

The main PMT features to measure are dark current (DC) and gain. In order to calculate the gain, a controlled amount of light, at the single-photoelectron (SPE) level, is driven inside the dewar using an optical fiber and a diffuser from Thorlabs\footnote{https://www.thorlabs.com/}. The light source used for the gain measurement is a PicoQuant GmbH\footnote{https://www.picoquant.com/} laser head, with a 405 nm wavelength where the PMT quantum efficiency is maximal. The PMT output is measured with a V965A CAEN\footnote{http://www.caen.it/} Charge-to-Digital Converter (QDC) in a 200\,ns window. The QDC integration window and the light source are synchronized using a dual output signal generator. To perform the DC acquisition, a V895 discriminator and a V560E scaler tandem from CAEN is used. The PMTs are biased using a CAEN's N1470 power supply. The DAQ and LN$_2$ filling control are implemented with LabVIEW software\footnote{http://www.ni.com/}.

To characterize the gain of each PMT, the SPE spectrum is fitted to two Gaussian functions, one modeling the pedestal and another the SPE peak. In Ref.~\cite{protoDUNEPMTs} a typical SPE spectrum from these PMTs is shown. The gain vs high voltage (HV), known as gain-voltage curve, is measured in 100\,V steps. Then, a fit is done following the power law $G= A V^B$ being $A$ and $B$ constants dependent on the number, structure, and material of the dynodes~\cite{ham2}. For the same HV, the gain at CT is lower than at RT.

The DC is the response of the PMT in absence of light. It is estimated as the average rate of detected signals larger than 7\,mV, which assures SPE triggering at an operating gain of 10$^7$. It is known that the main contribution to the DC at RT is the thermionic emission~\cite{ham2}. However, a non-thermal contribution increases the DC rate at CT \cite{meyer2,meyer3}. With a temperature change, the DC undergoes two effects: it decreases with temperature but increases with gain. As the gain increases with temperature the decrease of the DC with temperature is compensated to some extent.

Measurements are taken with each PMT at CT during approximately one week. The first day (Day~1), the PMT is installed at the vessel filled with LN$_2$. The PMT needs around three days for thermalization (Days~2 to~4)~\cite{protoDUNEPMTs}. Then, gain and DC measurements are taken at atmospheric pressure (Day~5). The same day (Day~5), the pressure is increased from 1\,bar to 4\,bar in $\sim$30 minutes, and left overnight for PMT thermalization. On the following day (Day~6), gain and DC measurements at 4\,bar are taken, and the system is set again to 1\,bar. On the last day (Day~7) gain and DC measurements at 1\,bar are taken again to ensure no damage has been caused to the PMTs. As a PMT break or malfunctioning would be quickly detected, no more than a 24\,h test in over-pressure is required. Afterwards, the gain-voltage curve was taken at RT for all PMTs.

It has to be noted that LN$_2$ temperature at 1\,bar (atmospheric pressure) is 77\,K and at 4\,bar (3-bar over-pressure) is  91\,K. 

\section{Results}
\label{sec4}

\begin {figure}[ht]
\includegraphics[width=0.6\textwidth]{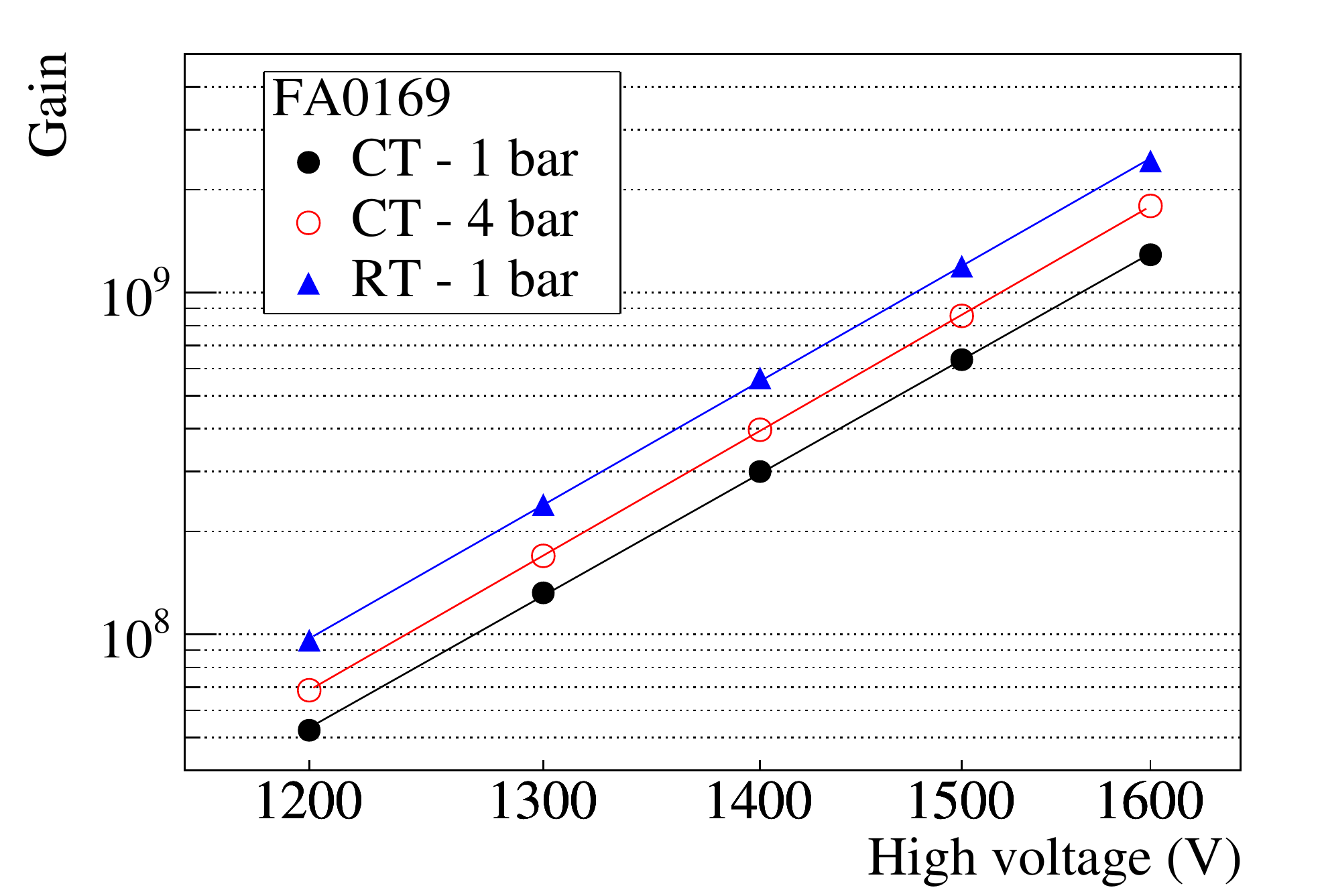}
\centering \caption{Gain vs HV at RT, and at two CT (77 K and 91 K) for the same PMT. The dots represent the measurements, and the lines the fit explained in section~\ref{sec3}. The gain dots include the error from the fit.}
\label{fig:gvshv}
\end {figure}

Gain-voltage curves are measured at each step, and the results for one PMT can be seen in Fig.~\ref{fig:gvshv}. For the same HV, the gain at CT is lower than at RT. Also at CT, the gain at 1\,bar is lower than at 4\,bar, due to the lower operating temperature. Being the operating gain 10$^8$ at RT (295\,K), on average for the three PMTs, the gain decreases by 46\,$\pm$\,1\% at CT at 1\,bar (77\,K), and by 21\,$\pm$\,8\% at 4\,bar (91\,K). Repetitive measurements on the same PMT gave consistent results.

\begin {figure}[ht]
\includegraphics[width=0.6\textwidth]{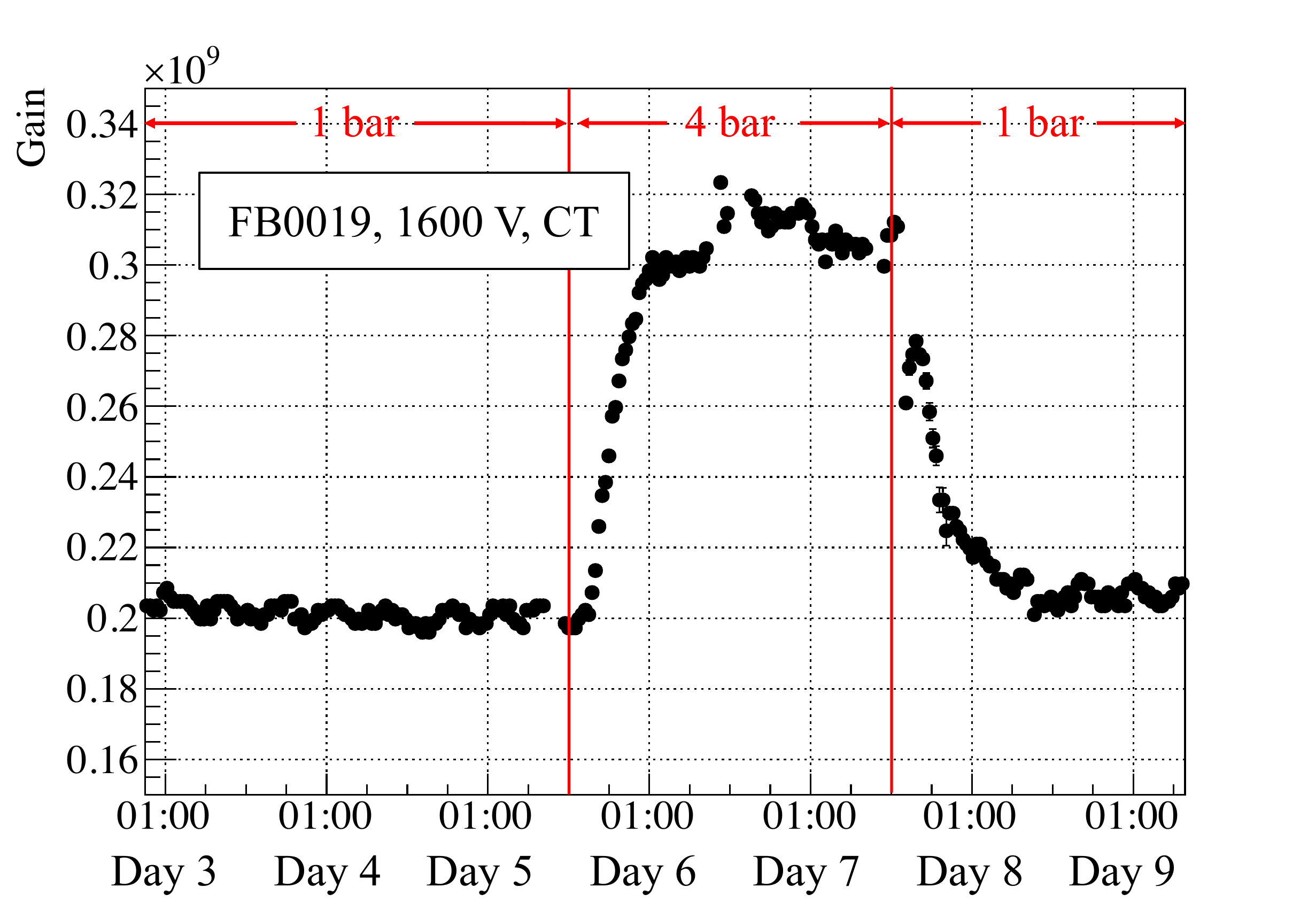}
\centering \caption{Gain evolution with time for one PMT. On Day 5 at 13:00\,h pressure is increased from 1 bar to 4 bar. On Day 7 at 13:00\,h pressure is decreased to 1 bar.}
\label{fig:gvst}
\end {figure}

The gain evolution when pressure is increased from 1\,bar to 4\,bar is shown in Fig.~\ref{fig:gvst} for one PMT. Gain increases with pressure because of the different operating temperatures and the same gain is recovered when the PMT is again at atmospheric pressure. It is observed that the gain stabilization takes $\sim$12\,h due to the PMT thermalization after the pressure is modified.

\begin {figure}[ht]
\includegraphics[width=0.6\textwidth]{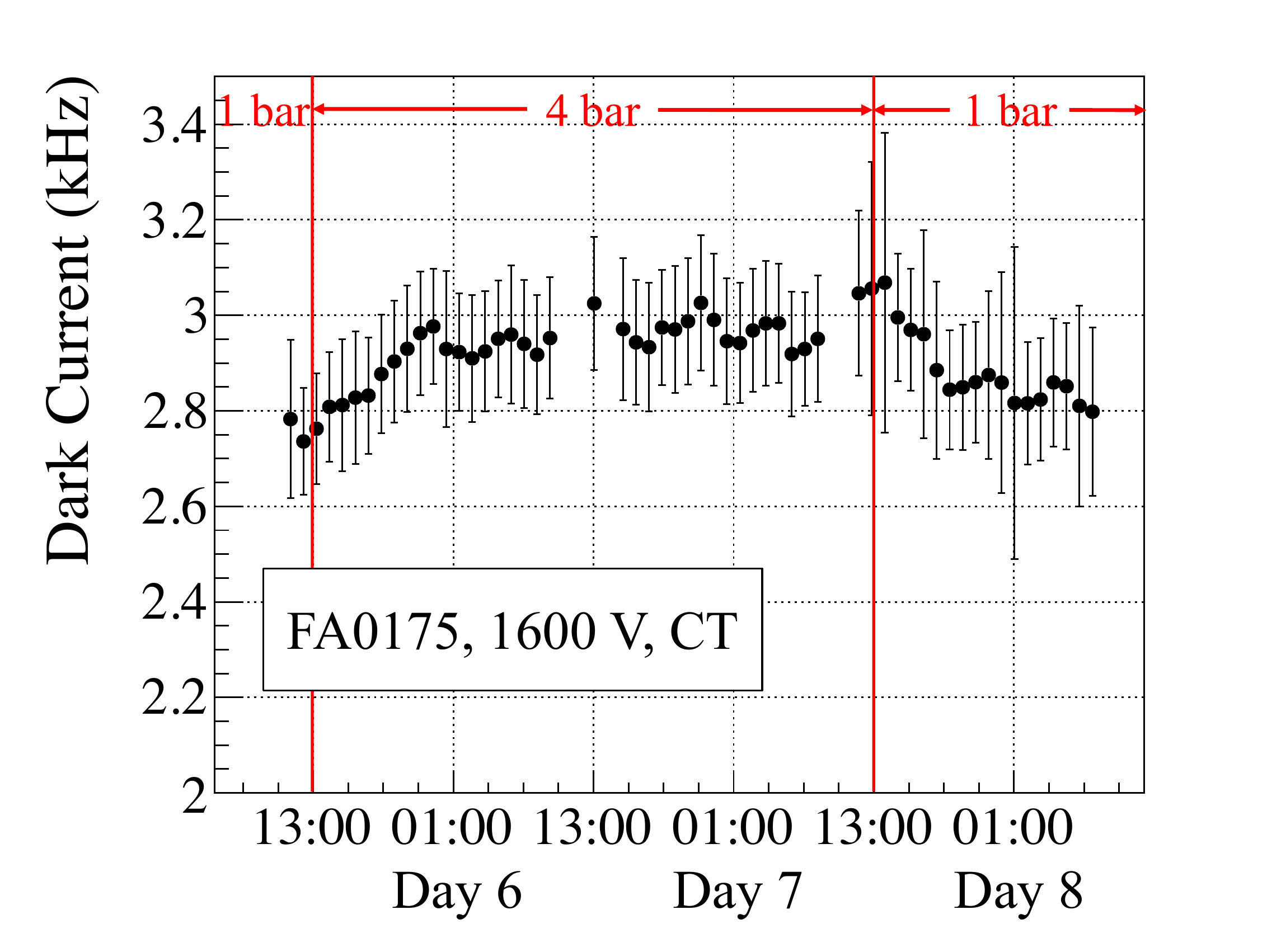}
\centering \caption{DC rate evolution with time for one PMT. On Day 5 at 13:00\,h pressure is increased from 1 bar to 4 bar. On Day 7 at 13:00\,h pressure is decreased to 1 bar.}
\label{fig:dc}
\end {figure}

Table~\ref{dc} shows the DC rate for a given voltage for all PMTs at 77 and 91\,K. As mentioned before, DC depends on temperature and gain, going both effects in opposite direction and no big change in the DC is expected. The evolution of the DC during the pressure cycle is shown in Fig.~\ref{fig:dc} for a PMT. No effect of the pressure change on the DC rate is observed, besides the one due to the gain and temperature changes, and no anomalies are found. Also in this case, $\sim$12\,h are needed for the DC rate to stabilize after a temperature change. 

\begin{table}[ht]
\centering
\caption{Dark current at a given HV.}
\begin{tabular}{cccccc}
\hline \hline
PMT  & P     & T    & HV     & G & DC	 \\
	 & (bar) & (K)	& (V)	& 	& (kHz) \\
\hline
FB0019 & 1 & 77 & 1600 & 2.0$\cdot$10$^8$ & 2.93\,$\pm$\,0.14 \\
FB0019 & 4 & 91 & 1600 & 3.0$\cdot$10$^8$ & 2.53\,$\pm$,0.15\\
FA0169 & 1 & 77 & 1500 & 6.4$\cdot$10$^8$ & 4.18\,$\pm$\,0.15 \\
FA0169 & 4 & 91 & 1500 & 8.6$\cdot$10$^8$ & 4.10\,$\pm$\,0.17\\
FA0175 & 1 & 77 & 1600 & 6.9$\cdot$10$^7$ & 2.78\,$\pm$\,0.13\\
FA0175 & 4 & 91 & 1600 & 1.1$\cdot$10$^8$ & 3.02\,$\pm$\,0.14 \\
\hline \hline
\end{tabular}
\label{dc}
\end{table}

None of the four PMTs showed any mechanical change after the over-pressure testing at CT and all three tested PMTs showed similar performance according to expectations.

\section{Conclusion}

As required for DUNE, Hamamatsu R5912-02mod PMTs support an absolute pressure of 4-bar without showing any mechanical or electrical damage at CT. The differences observed in the behavior are expected due to the change in the operating temperature. Hence, in terms of operating pressure, the Hamamatsu R5912-02Mod PMT model is validated for its use in the DUNE dual-phase module and in any other cryogenic detector operating at up to 4-bar absolute pressure.

\section*{Acknowledgment}

This project has received funding from the European Union Horizon~2020 Research and Innovation programme under Grant Agreement no.~654168; from the Spanish Ministerio de Economia y Competitividad (SEIDI-MINECO) under Grants no.~FPA2016-77347-C2-1-P, and MdM-2015-0509; from the Comunidad de Madrid; and the support of a fellowship from ''la Caixa" Foundation (ID 100010434) with code LCF/BQ/DI18/11660043. The authors thank Hamamatsu Photonics K. K. for their collaboration that was essential to perform this work.

\bibliographystyle{JHEP}
\bibliography{biblio}

% Please avoid comments such as "For a review'', "For some examples",
% "and references therein" or move them in the text. In general,
% please leave only references in the bibliography and move all
% accessory text in footnotes.

% Also, please have only one work for each \bibitem.

\end{document}